\documentclass[aps,prl,twocolumn,superscriptaddress]{revtex4}
\usepackage{epsfig}
\usepackage{amsmath}
\usepackage{subfigure}
\usepackage{wrapfig}
\usepackage{graphicx}
\usepackage{amssymb}
\usepackage{ulem}

\begin{document}

\title{Effective classical correspondence of the Mott transition}

\author{Danqing Hu}
\author{Jian-Jun Dong}
\affiliation{Beijing National Laboratory for Condensed Matter Physics and Institute
of Physics, Chinese Academy of Sciences, Beijing 100190, China}
\affiliation{School of Physical Sciences, University of Chinese Academy of Sciences, Beijing 100190, China}
\author{Li Huang}
\affiliation{Science and Technology on Surface Physics and Chemistry Laboratory, P.O. Box 9-35, Jiangyou 621908, China}
\author{Lei Wang}
\email[]{wanglei@iphy.ac.cn}
\author{Yi-feng Yang}
\email[]{yifeng@iphy.ac.cn}
\affiliation{Beijing National Laboratory for Condensed Matter Physics and Institute
of Physics, Chinese Academy of Sciences, Beijing 100190, China}
\affiliation{School of Physical Sciences, University of Chinese Academy of Sciences, Beijing 100190, China}
\affiliation{Songshan Lake Materials Laboratory, Dongguan, Guangdong 523808, China}

\date{\today}

\begin{abstract}
We derive an effective classical model to describe the Mott transition of the half-filled one-band Hubbard model in the framework of the dynamical mean-field theory with hybridization expansion of the continuous time quantum Monte Carlo. We find a simple two-body interaction of exponential form and reveal a classical correspondence of the Mott transition driven by a logarithmically divergent interaction  length. Our work provides an alternative angle to view the Mott physics and suggests a renewed possibility to extend the application of the quantum-to-classical mapping in understanding condensed matter physics.
\end{abstract}

\maketitle

The Mott transition is arguably one of the most fundamental concepts of correlated electrons and has been extensively investigated during the past decades \cite{Mott1949,Mott1961,Hubbard1964a,Hubbard1964b,Brinkman1970,Castellani1979,Georges1992,Rozenberg1992,Rozenberg1994,Imada1998,Kotliar1999,Kotliar2000}. In contrast to the band insulator at integer filling, the Mott insulator occurs at half integer filling due to strong onsite Coulomb interactions. It is beyond the conventional band picture and provides a basis for our understanding of many exotic properties in transition metal oxides including cuprates and manganese. In the framework of the dynamical mean-field theory (DMFT), the Mott transition is predicted to be a first-order transition with a coexisting insulating and metallic regime below the critical end point \cite{Georges1992,Rozenberg1992,Rozenberg1994,Imada1998,Caffarel1994,Majumdar1995,Rozenberg1995,Georges1996,Kotliar1999,Kotliar2000}. This has been confirmed experimentally in transition metal oxides such as V$_{2-x}$Ti$_x$O$_3$ \cite{McWhan1971}, in which a systematic analysis of the conductivity has revealed a scaling behavior near the critical end point resembling that of the liquid-gas transition \cite{Limelette2003}. It is therefore natural to ask if and how the Mott system can be mapped to a classical liquid-gas system.

The quantum-to-classical mapping has made important contributions in the history of condensed matter physics
\cite{Anderson1969,Anderson1970a,Anderson1970b,Schotte1970,Anderson1970c,Blume1970,Suzuki1971,Suzuki1976,Emery1974,Chui1975,Chakravarty1982,Spohn1985,Fannes1988,Wang1993,Novais2002,Shah2003}. The mapping is a generic property of quantum statistical mechanics, but often limited by a complex series expansion of the partition function. We show that such a situation may be improved with the help of a lately-developed machine learning approach. The latter has led to rapid progresses \cite{Carleo2019} in identifying phase transitions \cite{Wang2016,Carrasquilla2017,Broecker2017,Nieuwenburg2017,Hsu2018,Zhang2018,Zhang2019}, constructing many-body ground states \cite{Carleo2017,Choo2018,Glasser2018,Levine2019}, speeding up quantum Monte Carlo simulations \cite{Huang2017a, Huang2017b, Liu2017,Chen2018}, and optimizing tensor networks \cite{Guo2018,Han2018,Liao2019}. In this work, we explore the possibility of using the machine learning technique to construct a classical correspondence of the Mott system within the DMFT framework. We find that the quantum model can be mapped to a classical molecular gas with an effective two-body potential of an exponential form. The Mott transition is then in correspondence with the classical liquid-gas transition tuned by the range of the inter-molecular interaction.

\begin{figure}[b]
\centering
\includegraphics[width=0.42\textwidth]{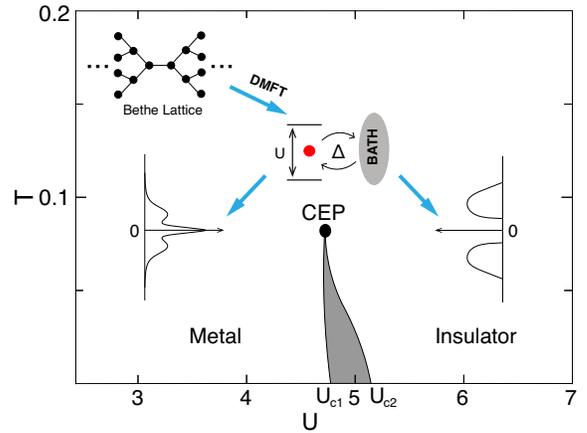}
\caption{A typical phase diagram of the half-filled one-band Hubbard model on the Bethe lattice. The point marks the critical end point (CEP) of the first-order Mott transition. $U_{c1}$ and $U_{c2}$ are the critical $U$ at zero temperature from the insulating and metallic sides, respectively. The shadow area is the coexisting (hysteresis) region of the two phases. The inset shows the structure of the Bethe lattice and its DMFT mapping to an impurity model coupled to a bath. The densities of states of the two phases are also plotted for comparison.}
\label{fig1}
\end{figure}

For simplicity, we discuss the Mott transition based on the one-band Hubbard model at half filling,
\begin{equation}
H=-t\sum_{\langle ij\rangle\sigma}\left(c^\dagger_{i\sigma}c_{j\sigma}+\text{H.c.}\right)+U\sum_i (n_{i\uparrow}-\frac12)(n_{i\downarrow}-\frac12),
\end{equation}
where $t$ is the hopping integral and $U$ is the onsite Coulomb interaction. In the framework of DMFT, the lattice problem is mapped to an impurity problem coupled with a self-consistent bath, as illustrated in the inset of Fig.~\ref{fig1}. Below we set $t=1$ as the energy unit and consider for simplicity only the paramagnetic phase and the Bethe lattice with a semicircular density of states,  $\rho_0(\epsilon)=\frac{1}{2\pi t}\sqrt{4t^2-\epsilon^2}$ \cite{Georges1996}. Our conclusions have been examined in some other lattices and found to be qualitatively unchanged. We then solve the impurity model using the continuous time quantum Monte Carlo method with hybridization expansion (CT-HYB) \cite{Werner2006, Gull2011}. The phase diagram of the Bethe lattice is sketched in Fig.~\ref{fig1}, where the shadow area is the coexisting (hysteresis) region for the first-order Mott transition below the critical end point (CEP). $U_{c1}$ ($U_{c2}$) is the critical values of $U$ for calculations starting from the insulating (metallic) side, which may vary for different lattices such as the bipartite lattice. The electron densities of states in both phases are also illustrated for comparison.

\begin{figure}[t]
\begin{center}
\includegraphics[width=0.42\textwidth]{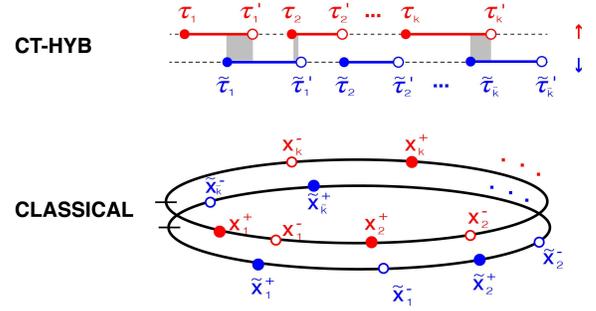}
\caption{Mapping between the CT-HYB configuration and the classical configuration of charged particles. For CT-HYB, the full and empty circles at the imaginary times $\tau_i$ ($\tilde{\tau}_i$) and $\tau_i^\prime$ ($\tilde{\tau}_i^\prime$) represent the creation and annihilation operators of spin up (down), respectively. The shaded areas indicate the overlap between the segments (solid lines) of two spin channels. In the classic model, the circles represent charged particles on a unit circle at $x_i^{+}=\tau_{i}/\beta$, $x_{i}^{-}  = \tau_{i}^\prime/\beta$, $\tilde{x}_i^{+}=\tilde{\tau}_{i}/\beta$ and $\tilde{x}_{i}^{-}  = \tilde{\tau}_{i}^\prime/\beta$.}
\label{fig2}
\end{center}
\end{figure}

The partition function of the impurity model has the form,
\begin{equation}
Z\propto \sum^{\infty}_{k,\tilde{k}=0} \int^{\beta}_{0} \prod^{k}_{i=1} ( d\tau_i d\tau^\prime_i  )  \prod^{\tilde{k}}_{j=1} (d\tilde{\tau}_j d\tilde{\tau}^\prime_j) \omega_{loc} \omega_{hyb},
\end{equation}
where $\omega_{\rm loc}=e^{U(L_{\rm tot}/2-O_{\rm tot})}$ and $\omega_{\rm hyb}=|\text{det}M^{-1}(\Delta)|$ are the local and hybridization weights for the configuration, $\{\mathcal{C}_k: \tau_1, \tau_1^\prime; \cdots; \tau_k, \tau_k^\prime\}$ and $\{\mathcal{C}_{\tilde k}: \tilde{\tau}_1,\tilde{\tau}_1^\prime; \cdots; \tilde{\tau}_{\tilde{k}},\tilde{\tau}_{\tilde{k}}^\prime\}$ of the two spin channels. $k$ and $\tilde{k}$ are their orders of expansion, respectively. In CT-HYB, as shown in Fig.~\ref{fig2}, $L_{\rm tot}$ is the sum of the lengths of all segments (solid line) and $O_{\rm tot}$ accounts for the overlap (shaded area) between segments of different spins. $M^{-1}(\Delta)$ is a $(k+\tilde{k}) \times (k+\tilde{k})$ matrix of the hybridization function $\Delta$ in imaginary time. Absent spin flip, the two spin channels are separated such that $\omega_{\rm hyb}=\omega_{\rm hyb}^{\uparrow}(\mathcal{C}_k)\omega_{\rm hyb}^\downarrow(\mathcal{C}_{\tilde{k}})$.  We have chosen CT-HYB rather than the interaction expansion (CT-INT) as the impurity solver \cite{Huang2017b}. In CT-INT, each configuration $\{\mathcal{C}_k: \tau_1, \cdots, \tau_k\}$ corresponds to a collection of interaction vertices, which can be mapped to only one type of classical particles and thus require three-body interactions to prevent the collapse of classical simulations \cite{Huang2017b}. By contrast, CT-HYB contains both creation and annihilation operators, which may be regarded as charged particles in the classical model and allow us to consider only the two-body interaction.

We first focus on the hybridization part, $\omega_{\rm hyb}$. The local part, $\omega_{\rm loc}$, is given by the total length of all segments and their overlaps as shown in Fig.~\ref{fig2}. Its logarithm can be directly expressed as a linear function of the coordinates ($x_i$) of the classical particles and hence has already a classical form, while $\omega_{\rm hyb}$ is more complicated and can only adopt a simple form through the mapping. Because of the spin symmetry, the two spin channels follow the same probability distribution function. We will only discuss one spin channel and  later consider their combination through $\omega_{\rm loc}$. As shown in Fig.~\ref{fig2}, if we regard the creation (annihilation) operators at the imaginary time $\tau_i$ ($\tau_i^\prime$) as a positive (negative) charge $q_i=+$ ($-$), the CT-HYB configurations are mapped to an ensemble of charged particles. Since the CT-HYB configuration is defined on a periodic imaginary time space with length $\beta$ (the inverse temperature) following the standard theory of quantum statistics, it is possible to restrict the classical particles on a unit circle with the classical coordinates: $x_i^{+}=\tau_{i}/\beta$, $x_{i}^{-}  = \tau_{i}^\prime/\beta$, $\tilde{x}_i^{+}=\tilde{\tau}_{i}/\beta$ and $\tilde{x}_{i}^{-}  = \tilde{\tau}_{i}^\prime/\beta$. The question is if a classical model may be constructed to reproduce the quantum weight of each configuration. 

For this, we first make the simplest assumption of two-body interactions and propose an energy function,
\begin{equation}
E_{\rm eff}(\mathcal{C}_k)=-\frac{1}{2}\textstyle\sum'_{i,q_i;j,q_j}V_{q_iq_j}(x_j^{q_j} - x_i^{q_i})+\mu_{\rm eff} k+E_0,
\label{Eff}
\end{equation}
where $V_{q_iq_j}(x_j^{q_j} - x_i^{q_i})$ is the two-body potential depending on the charges and distance of two different classical particles, $E_0$ is a constant background energy, and $\mu_{\rm eff}$ is an effective chemical potential of the classical model coupled to the number $k$ of charged pairs, which should not be confused with the usual chemical potential of the Hubbard model. $\sum'$ indicates that the self-interaction is excluded in the sum. To determine the exact form of the potential function, we use the Legendre expansion,
$V_{q_i q_j}(x) = \sum_{l=0} V_{q_i q_j}^l P_l(2x-1)$,
where $0\le x\le 1$ and $P_l(x)$ is the $l$-th order Legendre polynomial. The classical model is then fully determined by the parameters $V_{q_iq_j}^l$ (up to a sufficiently large cutoff of $l$) and $\mu_{\rm eff}$, which can be trained with linear regression by requiring $-E_{\rm eff}(\mathcal{C}_k)$ to match $\ln \omega_{\rm hyb}^\sigma(\mathcal{C}_k)$ for all tested CT-HYB configurations. We have collected 250 000 samples for each set of parameters. A single CT-HYB sample is picked up every 100 update steps during an additional iteration after the DMFT convergence. We then take the log-weights as the regressing target and apply the ridge regression with $L_2$ regularization of the strength $\alpha=10^{-3}$ to prevent overfitting. The machine learning approach minimizes the penalized residual sum of squares, $\left \| {\rm X}\omega-y \right \|^2_2+\alpha \left \| \omega\right \|^2_2$, with the coefficients vector $\omega$, the input configuration dependent matrix $X$ and the target value $y$. In our calculations, $\omega=(V_{qq^\prime}^0, \dots, V_{qq^\prime}^{l_m-1}, \mu_{\rm eff}, E_0)^T$ is the coefficient vector for the $l$-th order Legendre polynomial cut off at $l_m=30$. $X$ is an $N_0\times (4l_m+2)$ matrix determined by the energy function, $N_0=250\ 000$ is the size of the training samples, and $y$ is a vector that contains the log-weight of each sample configuration.

\begin{figure}[t]
\begin{center}
 \includegraphics[width=0.48\textwidth]{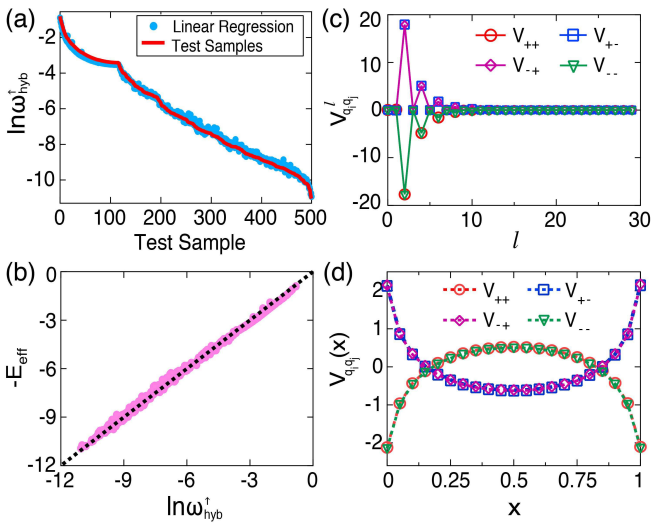}
 \caption{Construction of the classical model for CT-HYB configurations. (a) Comparison of the log-weight of test samples and the linear regression results. (b) Comparison of the hybridization log-weight and the effective classical energy, $-E_{\rm eff}$. Each point represents a test sample and the dotted line is a guide to the eye. (c) The derived Legendre coefficients $V_{q_i q_j}^{l}$ for the two-body interactions. (d) The two-body interactions $V_{q_i q_j}(x)$ as functions of the distance $x$ on the unit circle. The parameters are $U=3$ and $\beta = 30$.}
\label{fig3}
\end{center}
\end{figure}

Figure~\ref{fig3} gives the typical results for the Bethe lattice with the Coulomb interaction $U=3$ and the inverse temperature $\beta=30$. As shown in Figs.~\ref{fig3}(a) and \ref{fig3}(b), we obtain an excellent agreement between the effective energy function and the calculated log-weight. The proposed classical model indeed captures the quantum distribution of the original Hubbard Hamiltonian. The deviation of the energy fit can be measured using $\sigma=N^{-1}\sum_i(y_{i,{\rm fit}}-y_i)^2$, where $N$ is the size of the test samples and set to 500 in our calculations. We find that $\sigma$ has the value of about 0.01 and shows no peculiar (singular) change over the whole phase diagram. The Legendre coefficients $V_{q_iq_j}^l$ are plotted in Fig.~\ref{fig3}(c). The rapid decay of their magnitude for large $l$ confirms the validity of the expansion, even at the critical end point. Figure~\ref{fig3}(d) compares the derived two-body interactions, $V_{q_i q_j}(x)$, as functions of distance. The plot reveals a number of interesting symmetries in $V_{q_iq_j}$, which may be rationalized as follows. First, for a classical model, one typically expects $V_{q_i q_j}(x)=V_{q_j q_i}(x)=V_{q_j q_i}(1-x)$ between any two particles on the unit circle. Second, inserting a segment whose length approaches zero should not change the total energy. In the quantum Monte Carlo simulations, this operation does not generate a new configuration and is therefore typically not considered. However, in the classical model, this corresponds to the situation of adding a pair of particles of opposite charges at the same location, whose interactions with any third particle should always cancel. The latter implies a generic rule of the classical interactions between charged particles, namely, $V_{++}=V_{--}=-V_{-+}=-V_{+-}=-q_iq_jV(x)$. Third, it immediately follows from the above constraints that the interaction $V(x)$ must be symmetric with respect to $x=1/2$, namely, $V(x)=V(1-x)$ for $0\le x\le 1$.

\begin{figure}
\centering
\includegraphics[width=0.48\textwidth]{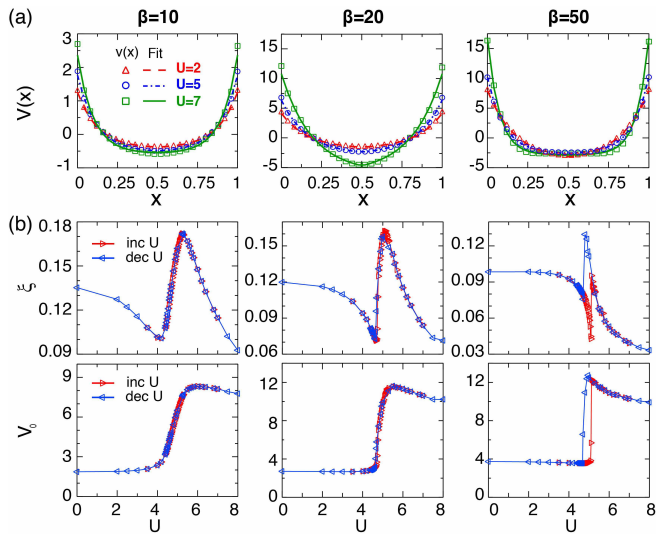}
\caption{(a) Exponential fit of the effective potential for different values of $U$ and $\beta$, showing excellent agreement in all three regimes. (b) The fitting values of $V_0$ and $\xi$ in the classical model after DMFT convergence with gradually increasing ($\rhd$) or decreasing ($\lhd$) $U$.}
\label{fig4}
\end{figure}

The functional form of $V(x)$ thus provides key information on the classical particle system. Quite unexpectedly, as shown in Fig.~\ref{fig4}(a), we find that it can be well fitted with an exponential function for all parameter regimes of $U$ and $\beta$ regardless of the metallic or insulating phases:
\begin{equation}
V(x)=V_0e^{-\min\{x, 1-x\}/\xi}+V_1,
\end{equation}
where $V_0$ and $V_1$ are both constants and $\xi$ reflects an effective range of the two-body interaction. This suggests that the simple  form captures the essential physics of the hybridization function. Since we are dealing with the self-consistent bath coupled to the impurity, the exponential decay seems to imply that the lattice effect is to screen the local correlation on a finite length (or imaginary time) scale $\xi$. It will be interesting to see if such a form may still be valid in more general cases such as away from half filling or with multiple orbitals. Our analyses seem independent of these details as long as the model does not contain any spin or orbital mixing.

The above potential can be further simplified by removing the constant term $V_1$. Since the charged particles always appear pairwise in CT-HYB configurations, the overall effect of $V_1$ on the total energy is nothing but a term, $-V_1 k$, which can be absorbed by redefining $(\mu_{\rm eff} - V_1) \rightarrow \mu_{\rm eff}$ in Eq.~(\ref{Eff}). Moreover, imagining that we insert again $N$ segments of almost zero length, the variation of the total energy should also be zero, namely $\delta E_{eff}=-V_0 N +\mu_{\rm eff} N=0$. This requires $\mu_{\rm eff}=V_0$, which was first found unexpectedly in all our fittings. We have thus only two parameters in the effective model and the total energy function adopts a very simple approximate form,
\begin{equation}
E_{\rm eff}(\mathcal{C}_k)-E_0=\frac{V_0}{2}\sum_{i,q_i;\,j,q_j}q_iq_j e^{-\delta_{q_jq_i}/\xi},
\label{Eeff}
\end{equation}
where the $k$-term is absorbed as a self-interaction for $i=j$ and $q_i=q_j$ and the distance between two charged particles is defined as, $\delta_{q_jq_i}=\min\{|x_j^{q_j}-x_i^{q_i}|, 1-|x_j^{q_j}-x_i^{q_i}|\}$, including self-interaction. This energy function incorporates faithfully the major effect of the lattice correlations on the local impurity model in the DMFT iterations. For examination, we have inserted it back into the Monte Carlo simulations after the parameters were determined with a small set of training samples, the results agree well with the pure quantum simulations.

Thus all informations on the Mott transition are squeezed into the interaction potential, $V_0$, and its effective range, $\xi$, in the classical model. To study how these two parameters behave across the Mott transition, we plot in Fig.~\ref{fig4}(b) their variations as a function of $U$ for  $\beta=$10, 20, 50, which are above, near and below the Mott critical end point, respectively. The results were obtained by gradually increasing $U$ from $U=0$ on the metallic side to $U=8$ on the insulating side and then annealing back to the metallic state.  For each $U$, we took the input from the converged solution of previous $U$ at the same $\beta$. As expected, we see a jump and a clear hysteresis at $\beta=50$ in both parameters due to the first-order nature of the Mott transition. For $\beta=10$ above the Mott critical end point, the jump turns into a smooth crossover and both parameters vary continuously with $U$.

\begin{figure}
\centering
\includegraphics[width=0.48\textwidth]{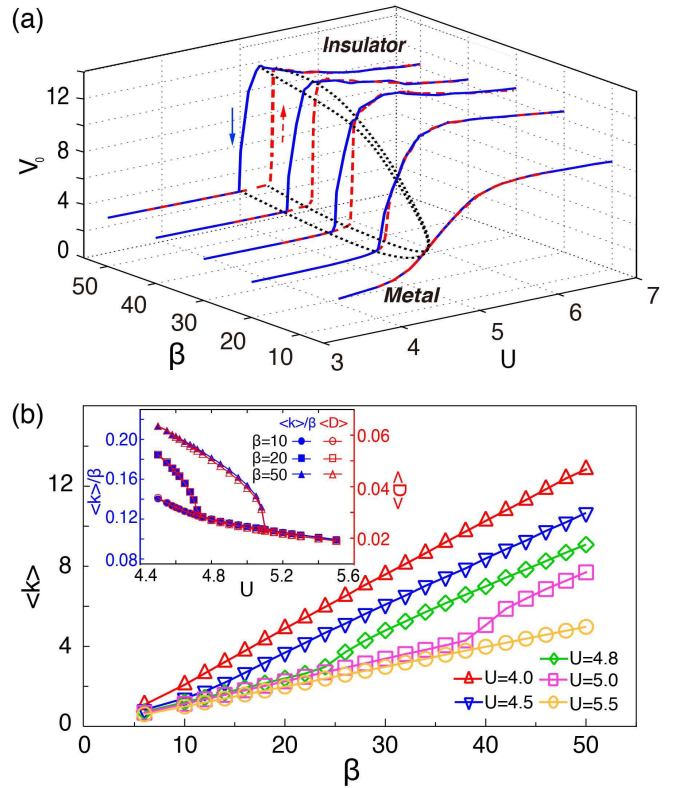}
\caption{(a) Variations of $V_0$ as functions of $\beta$ and $U$, showing a rapid change across the Mott transition and a slight change in other parameter regimes. The arrows indicate the direction of decreasing (solid line) or increasing (dashed line) $U$ for DMFT calculations. The dotted lines mark the region of numerical hysteresis of the first-order transition. (b) The pair number, $\langle k\rangle$, as a function of the inverse temperature $\beta$ for different values of $U$, showing the slope change across the Mott transition. The inset compares the calculated pair density, $\langle k\rangle/\beta$, and the double occupancy, $\langle D\rangle$, for increasing $U$.}
\label{fig5}
\end{figure}

The overall variation of $V_0$ with respect to the original parameters $U$ and $\beta$ is summarized in Fig.~\ref{fig5}(a). Surprisingly, we see $V_0$ only undergoes a rapid change of roughly the factor of 3 across the transition and otherwise varies only slightly with $U$ and $\beta$. This is reminiscent of the volume change in the classical liquid-gas transition. $V_0$ thus plays a passive role in the phase transition of the classical model. Interestingly, as shown in Fig.~\ref{fig5}(b), the particle number $\langle k\rangle$ as a function of $\beta$ changes its slope across the transition, which is roughly 0.3 in the metallic state but decreases to 0.1 in the insulating state, just the opposite of that of $V_0$. In addition, the pair density, $\langle k\rangle/\beta$, has a linear relationship with the double occupancy, $\langle D\rangle$, around the critical $U$ (see the inset). Thus the variation of $\langle k\rangle$ indeed reflects the situation of local electron occupations and contains some information of the Mott transition.

\begin{figure}
\centering
\includegraphics[width=0.48\textwidth]{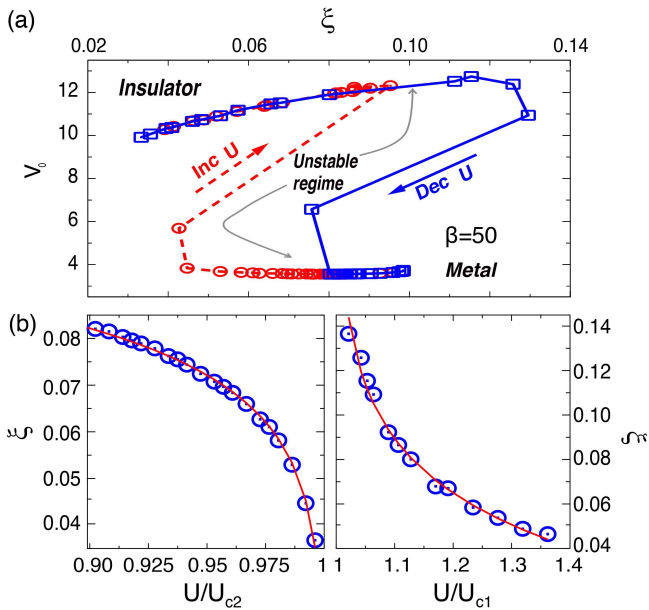}
\caption{ (a) Illustration of the parameter flow on the $V_0-\xi$ space with increasing (circle) or decreasing (square) $U$ at $\beta = 50$, showing almost constant $V_0$ and an instablity with varying $\xi$. (b) Logarithmic fit (solid lines) of the effective length, $\xi$, from both sides of the Mott transition for $\beta = 50$. $U_{c2}$ and $U_{c1}$ are the critical values of $U$ from the metallic and insulating side, respectively.}
\label{fig6}
\end{figure}

It is therefore speculated that $\xi$ is the primary driving force for the phase transition. Figure~\ref{fig6}(a) plots the parameter flow in the $V_0-\xi$ space for both increasing and decreasing $U$ at $\beta=50$. While $V_0$ changes only slightly in each phase, we see $\xi$ varies greatly until a sudden transition is triggered. Since $\xi$ represents the effective range of the two-body interaction, the transition is therefore controlled by the scattering length between two charged particles. As a matter of fact, as shown in Fig.~\ref{fig6}(b), we find a logarithmic dependence, $\xi \sim \pm\ln|1-U/U_c|$ from either side of the phase diagram. This indicates that the effective interaction becomes progressively short or long range approaching the Mott transition. From the metallic side, the potential turns gradually local as $U\rightarrow U_{c2}$ and the particles become asymptotically free before they expand into the gas state; while starting from the insulating side, $\xi$ becomes gradually divergent as $U\rightarrow U_{c1}$ and the particles get more and more correlated until they eventually condense into a liquid. It is thus conceivable that the Mott metal-to-insulator transition corresponds roughly to a classical liquid-to-gas transition controlled by the single parameter $\xi$. It is well known that a one-dimensional classical system with short-range interactions may not have a positive temperature phase transition \cite{Cuesta2004}. A well known example is the one-dimensional Ising model with the nearest-neighbor interaction \cite{Ising1925}. However, a phase transition can occur once long-range interactions are introduced \cite{Frohlich1982}.

The classical correspondence provides to some extent an alternative angle to view the spin and charge dynamics of the Mott physics. One should combine the two spin channels through the local weight $\omega_{\text loc}$ into a single circle. Since $U$ is large, all the segments are intended to cover together the whole circle to reduce the Coulomb energy. Thus each charged particle is roughly bound to a nearby particle of opposite spin and charge. Then the CT-HYB configurations are reduced to a single circle and the corresponding classical model turns into a gas of diatomic molecules with an additional intra-molecular potential $V_{\rm m}(x)\propto \sum_{i,q_i}|x_{i\uparrow}^{q_i}-x_{i\downarrow}^{-q_i}|$. The charge and spin fluctuations correspond to the internal and global configurations of the molecular gas, respectively. We should note that despite the good agreement of the energy function, the classical model should by no means be viewed as an exact representation of the quantum model. It might not be able to reproduce certain subtle properties of the original model, in particular the long-time dynamics in the very vicinity of the critical end point. Nevertheless, its simple and explicit form may still capture some truth of the underlying physics. Considering that the classical mapping of the Kondo problem has led to the idea of the poor man's scaling \cite{Anderson1970c}, it will be interesting to see if future work might reveal some deeper structure of the Mott transition based on the derived classical correspondence here. For numerical studies, our energy function may be applied as a reference system for the quantum simulations. As discussed in Ref. \cite{Huang2017b}, once the parameters are obtained with a relatively small number of test configurations, the Monte Carlo simulations may be accelerated by an improved acceptance probability $p(\mathcal{C}\rightarrow \mathcal{C}^\prime)=\min\left\{1, \frac{e^{-E_{\rm eff}(\mathcal{C})}}{e^{-E_{\rm eff}(\mathcal{C}^\prime)}}\frac{\omega(\mathcal{C}^\prime)}{\omega(\mathcal{C})}\right\}$.

To summarize, we have constructed an effective classical correspondence for understanding the Mott transition of the half-filled Hubbard model in the framework of DMFT with the CT-HYB algorithm. The derived model may be regarded as a charged molecular gas with a simple two-body interaction of the exponential form. Our analysis suggests that the Mott transition may correspond to a classical liquid-gas transition of the molecules driven by the effective range of the particle interaction. The correlation length exhibits logarithmic behaviors approaching the transition. It should be noted that the above picture was derived in the framework of DMFT. For an exact lattice treatment, the quantum model in $d$ spatial dimension corresponds to a classical model in $d+1$ dimension. Then one may ask if similar two-body interactions will still be valid, possibly with anisotropic correlation lengths, $\xi^{\mu}_{\rm eff}$, whose values reflect the screening of the Coulomb potential along different spatial or temporal axes. A verification of this scenario demands numerical simulations of the lattice model using more sophisticated approaches.

This work was supported by the National Natural Science Foundation of China (11974397, 11874329) and the National Key Research and Development Program of China (2017YFA0303103).

\end{document}